\documentstyle[12pt]{article}
\newlength{\pubnumber} 

\catcode`\@=11
\@addtoreset{equation}{section}
\def\section{\@startsection{section}{1}{\z@}{3.5ex plus 1ex minus .2ex}
 {2.3ex plus .2ex}{\large\bf}}
\def\subsection{\@startsection{subsection}{2}{\z@}{2.3ex plus .2ex}
 {2.3ex plus .2ex}{\bf}}

\begin{document}

\begin{titlepage}
\samepage{
\setcounter{page}{1}
\rightline{UFIFT-HEP-96-15, DPFD96/TH/28}
\rightline{\tt hep-th/9606063}
\rightline{June 1996. Revised November 1996.}
\rightline{To be published in Phys. Rev. Lett.}
\vfill
\begin{center}
 {\Large \bf  Duality of $x$ and $\psi$ \\
and a Statistical Interpretation of Space in Quantum Mechanics\\}
 {\large Alon E. Faraggi$^{1,2}$\footnote{
   E-mail address: faraggi@phys.ufl.edu}
   $\,$and$\,$ Marco Matone$^{3}$\footnote{
   E-mail address: matone@padova.infn.it}\\}
\vspace{.1in}
 {\it $^{1}$   Institute for Fundamental Theory, Department of Physics, \\
        University of Florida, Gainesville, FL 32611,
        USA\footnote{Permanent address.}\\}
\vspace{.025in}
{\it $^{2}$  Theory Division, CERN, 
                1211 Geneva, Switzerland \\}
\vspace{.025in}
{\it $^{3}$   Department of Physics ``G. Galilei'' -- Istituto 
                Nazionale di Fisica Nucleare\\
        University of Padova, Via Marzolo, 8 -- 35131 Padova, Italy\\}
\end{center}
\vfill
\begin{abstract}
  {\rm
We introduce a ``prepotential'' ${\cal F}$ in quantum mechanics
and show that
the coordinate $x$ is proportional to the Legendre transform of
${\cal F}$ with respect to the probability density.
Inversion of the Schr\"odinger equation
leads us to consider a $x$--$\psi$ duality
related to a modular symmetry.
The scaling of $x$ is determined by
the ``beta--function'',
suggesting that in quantum mechanics the space coordinate
is a macroscopic variable of a statistical system
with $\hbar$ playing the role of scale.
The formalism is extended to higher
dimensions and to the Klein--Gordon equation.

}
\end{abstract}
\vfill
\smallskip}
\end{titlepage}

\setcounter{footnote}{0}

\def\beq{\begin{equation}}
\def\eeq{\end{equation}}
\def\beqn{\begin{eqnarray}}
\def\eeqn{\end{eqnarray}}
\def\AEF{A.E. Faraggi}
\def\NPB#1#2#3{{\it Nucl.\ Phys.}\/ {\bf B#1} (19#2) #3}
\def\PLB#1#2#3{{\it Phys.\ Lett.}\/ {\bf B#1} (19#2) #3}
\def\PRD#1#2#3{{\it Phys.\ Rev.}\/ {\bf D#1} (19#2) #3}
\def\PRL#1#2#3{{\it Phys.\ Rev.\ Lett.}\/ {\bf #1} (19#2) #3}
\def\PRT#1#2#3{{\it Phys.\ Rep.}\/ {\bf#1} (19#2) #3}
\def\MODA#1#2#3{{\it Mod.\ Phys.\ Lett.}\/ {\bf A#1} (19#2) #3}
\def\IJMP#1#2#3{{\it Int.\ J.\ Mod.\ Phys.}\/ {\bf A#1} (19#2) #3}
\def\nuvc#1#2#3{{\it Nuovo Cimento}\/ {\bf #1A} (#2) #3}
\def\etal{{\it et al,\/}\ }
\hyphenation{su-per-sym-met-ric non-su-per-sym-met-ric}
\hyphenation{space-time-super-sym-met-ric}
\hyphenation{mod-u-lar mod-u-lar--in-var-i-ant}


\setcounter{footnote}{0}
In the last couple of years Seiberg--Witten theory \cite{sw}
has shed new light
on some aspects of supersymmetric quantum field theories.
An important quantity in
this theory is the prepotential
${\cal F}$ as it fixes the low--energy dynamics.
In terms of ${\cal F}(\Phi)$, which is a holomorphic function
of the chiral superfield $\Phi$,
one can express the dual variable
$\Phi_D={\cal F}'(\Phi)$ and the effective coupling
constant $\tau ={\cal F}''(\Phi)$.
The quantum moduli space of the theory is parameterized
by the gauge invariant parameter $u=\langle{\rm Tr}\, \phi^2\rangle$,
where $\phi$ is the scalar component of $\Phi$.
In Seiberg--Witten theory a method has been developed to
invert the function $a=a(u)$ to $u=u(a)$ where
$a=\langle\phi\rangle$  \cite{m1}.
In this theory a second--order differential equation is
written down for the moduli parameters $a(u)$ and $a_D(u)$.
The prepotential enables the inversion procedure and allows
interesting interpretation of second--order differential equations.

Following these ideas,
we derive a method to invert the Schr\"odinger
wave--function $\psi=\psi(x)$ to $x=x(\psi)$.
We define a ``prepotential'' ${\cal F}$ as function of $\psi$
such that the dual variable
$\psi_D=\partial{\cal F}/\partial{\psi}$ is a solution
of the Schr\"odinger equation.
In this formalism
the quantum dynamics is described by ${\cal F}$,
which satisfies a 
non--linear third--order differential equation which replaces
the Schr\"odinger equation.
The inversion formula shows that $x$ is
the Legendre transform of ${\cal F}$ with respect
to the probability density, implying that in quantum mechanics
the space may be seen as
a macroscopic variable of a statistical system.
In this context we show that the scaling properties of $x$ with respect to
$\tau=\partial_\psi^2{\cal F}$ are determined in terms of the
``beta--function''
$\hbar \partial_\hbar \tau$.

Let us consider the
Schr\"odinger equation
\begin{equation}
\left(-{{\hbar^2}\over {2m}}\partial_x^2+V(x)\right)\psi=E\psi,
\label{erwin}
\end{equation}
where $E$ is in the physical spectrum of the Schr\"odinger operator.
In a general Schr\"odinger problem, such as Eq.(\ref{erwin}),
for each $E$ one can have one or two physical solutions.
Let $\psi_E$ denote a physical solution of Eq.(\ref{erwin}) and
${\psi_E}_D$ a solution of Eq.(\ref{erwin}) linearly independent from
$\psi_E$.
We define the prepotential ${\cal F}_E$ by
\begin{equation}
{\psi_{E}}_D={{\partial{\cal F}_E(\psi_E)}\over{\partial\psi_E}},
\label{psied}
\end{equation}
and consider
\begin{equation}
\partial_x{\cal F}_E={\psi_E}_D\partial_x \psi_E=
{1\over 2}\left[\partial_x(\psi_E{\psi_E}_D)+W\right],
\label{fx}
\end{equation}
where by Eq.(\ref{erwin}) the Wronskian
$W={\psi_E}_D\partial_x\psi_E-\psi_E\partial_x{\psi_E}_D$
is a constant.
The crucial point is that Eq.(\ref{fx}) can be integrated exactly to
\begin{equation}
{\cal F}_E={1\over 2}\psi_E{\psi_E}_D+
{W\over 2} x +{\rm c},
\label{f}\end{equation}
with ${\rm c}$ a constant which
by Eq.(\ref{psied})
we can set to 0.
 It is easy to check that Eq.(\ref{f}) is
equivalent to
\begin{eqnarray}
{\cal F}_E(\psi_E)=
&&\psi_E^2\left[{Wx_0+\psi_{E0}{\psi_E}_{D0}
\over 2\psi^2_{E0}}\right.\nonumber\\
&&\left.~~-W\int^{\psi_E}_{{\psi_E}_0}dy
{\cal G}_E(y)y^{-3}\right],
\label{iqwnd}\end{eqnarray}
where $\psi_{E0}\equiv \psi_{E}(x_0)$, ${\psi_E}_{D0}\equiv
{\psi_E}_{D}(x_0)$ and the notation $x={\cal G}_E(\psi_E)$ has been
introduced in order to denote the
functional dependence of $x$ on $\psi_E$.
By rescaling $\psi_E$ we can set $W=-{2\sqrt{2m}\over \hbar}$,
so that we have
\begin{equation}
{\sqrt{2m}\over \hbar}x(\psi_E)={1\over2}{\psi_E}{\partial
{\cal F}_E\over \partial \psi_E}-{\cal F}_E,
\label{iu1}\end{equation}
which we rewrite in the ``canonical form''
\begin{equation}
{\sqrt{2m}\over \hbar} x(\psi_E)={\psi^2_E}{\partial
{\cal F}_E\over \partial (\psi^2_E)}-{\cal F}_E,
\label{iu}\end{equation}
showing that the classical coordinate is proportional
to the Legendre transform
of the prepotential with respect to $\psi_E^2$.
Duality of the Legendre transform yields
\begin{equation}
{\hbar\over\sqrt{2m}}{\cal F}_E=\phi_E \partial_{\phi_E} x-x,
\label{uixk}\end{equation}
where $\phi_E=\partial_{(\psi_E^2)}{\cal F}_E={\psi_E}_D/2\psi_E$.
Therefore ${\cal F}_E$ is the Legendre transform of
${\sqrt{2m}\over \hbar}x$ and vice--versa.

In a quantum mechanical problem we can distinguish two cases depending on
whether $\psi_E$ and $\overline\psi_E$ are or are not linearly dependent
functions. In the former case (e.g. the harmonic oscillator)
the quantity $\psi_E^2$
is nothing else but the (unnormalized) probability density $\rho_E$.
Therefore, if $\overline \psi_E\propto\psi_E$,
Eq.(\ref{iu})
implies that the classical coordinate $x$ is proportional to the
Legendre transform of the prepotential with respect to $\rho_E$.
Furthermore, by Eqs.(\ref{iu1},\ref{uixk}) it follows that
\begin{equation}
\rho_E\equiv \psi_E^2={\sqrt{2m}\over \hbar}\partial_{\phi_E}x,
\label{news}\end{equation}
that is, $x$ is the generating function
for the probability density at $x$ itself.

Reality of the Schr\"odinger operator
implies that $\overline\psi_E$ is still a solution of Eq.(\ref{erwin}).
Therefore, if $\overline\psi_E\not\propto \psi_E$ (e.g. in the case of
the free particle, where
${\psi_E}_D=\overline \psi_E \propto
{\psi_E}^{-1}\propto e^{ipx/\hbar},~E=p^2/2m$),
then we can set\footnote{Note that with this choice $W$ is purely imaginary.
We can choose, without loss of generality, a normalization for $\psi_E$
itself such that
$W=-{2i\sqrt{2m}\over \hbar}$.}
\begin{equation}
{\psi_E}_D=\overline \psi_E,
\label{gdft}\end{equation}
and by (\ref{iu})
\begin{equation}
\rho_E\equiv |\psi_E|^2={i\sqrt{2m}\over \hbar}x+2{\cal F}_E,
\label{2}\end{equation}
showing that the probability density of finding the particle at
$x$ is proportional to $x$ itself with an additive
correction which is proportional to the prepotential.

${\cal F}$ plays a crucial role as it encodes the information
on the microscopic theory.
In particular
the Schr\"odinger equation can be replaced by ($'\equiv \partial_{\psi_E}$)
\begin{equation}
4{\cal F}_E'''+(V(x)-E))\left(
{\cal F}_E'-\psi_E{\cal
F}_E''\right)^3=0,
\label{ioc}\end{equation}
where $\hbar$ appears
only through $V(x)$ (with $x=x(\psi_E)$ given by (\ref{iu1})).
Eq.(\ref{ioc}) is obtained from Eqs.(\ref{erwin},\ref{iu1})
by following the method introduced
in \cite{m1,bonelli}. In particular, by inverting
the Schr\"odinger equation, we obtain
\begin{equation}
{\hbar^2\over 2m}\partial_{\psi_E}^2x=\psi_E(E-V(x))
(\partial_{\psi_E}x)^3,
\label{dualerwin}\end{equation}
which can be seen as dual to Eq.(\ref{erwin}).
These dual formulations of quantum mechanics may generate
different structures once one considers the second quantization or
alternatively quantize
the expansion of $x$ in powers of the wave--function.
In order to illustrate this point we first
consider the dual power expansions
\begin{equation}
\psi_E=\sum_j \alpha_j^E x^j \quad \Longleftrightarrow \quad
x=\sum_k \beta_k^E\psi_E^k,
\label{mirrosymmetry}\end{equation}
and note that their structure
suggests considering  the $x$--$\psi_E$ duality
as reminiscent of the ``mirror symmetry phenomenon''
first observed for Calabi--Yau threefolds.
We note that a similar remark was made in connection
with the differential equation (and its inverse)
satisfied by the generating function
for Weil--Petersson volumes
of moduli spaces of punctured Riemann spheres \cite{KaufmannManinZagier}.

Before considering the quantization of
(\ref{mirrosymmetry}),
it is worth noticing that the above
 structures are related to the modular symmetry
which underlies quantum mechanics. In particular,
the relation between the space coordinate, the prepotential and
the wave--function, is related to the
basic fact that any linear combination of $\psi_E$ and
${\psi_E}_D$ is still a solution of the Schr\"odinger equation.
For the same reason, the formalism is invariant under the
transformations
\begin{equation}
\tilde{\psi_E}_D=A\psi_E+B{\psi_E}_D, \qquad
{\tilde\psi}_E=C\psi_E+D{\psi_E}_D,
\label{kjd}\end{equation}
implying, in particular, that
Eqs.(\ref{iu1},\ref{ioc})
are modular invariant. This symmetry
can be also explicitly checked by using the transformation
properties of ${\cal F}_E$ which follow
by comparing $\tilde{{\psi}_E}_D=\partial \widetilde{\cal F}
(\tilde\psi_E)/\partial \tilde \psi_E$ with (\ref{kjd}),
\begin{eqnarray}
\delta {\cal F}_E&=&
{AC\over 2}{\psi_E}_D^2+
{BD\over 2} \psi_E^2+BC\psi_E{\psi_E}_D\nonumber\\
&=&{1\over 4}
v^t\left[G^t
\left(\begin{array}{c}0\\1
\end{array}\begin{array}{cc}1\\0\end{array}\right)G-
\left(\begin{array}{c}0\\1
\end{array}\begin{array}{cc}1\\0\end{array}\right)\right]v
\label{jfwfd}
\end{eqnarray}
where $\delta {\cal F}_E=
\widetilde {\cal F}_E(\tilde \psi_E)-{\cal F}_E(\psi_E)$,
 $G=\left(\begin{array}{c}A\\C\end{array}
\begin{array}{cc}B\\D\end{array}\right)\in SL(2,C)$ and
$v=\left(\begin{array}{c} {\psi_E}_D\\ {\psi_E}
\end{array}\right)$.

Let us now consider
the quantization of the expansions (\ref{mirrosymmetry}).
First note that we have the consistency conditions
\begin{eqnarray}
\psi_E&=&\sum_j\alpha_j^E\left(\sum_k\beta_k^E\psi_E^k\right)^j
\Longleftrightarrow~\nonumber\\
x&=&\sum_k \beta_k^E\left(\sum_j\alpha_j^E x^j\right)^k,
\label{mirrosymmetry2}
\end{eqnarray}
which imply an infinite set of
relations. Similar relations arise
also for an arbitrary
state described by a wave--function $\psi$. In particular,
expanding $\psi$ in a given basis
$\{\psi_j\}$,
Eq.(\ref{mirrosymmetry}) generalizes to
\begin{eqnarray}
\psi&=&\sum_j a_j\psi_j=\sum_j\alpha_j x^j \quad \Longleftrightarrow
\quad\nonumber\\
x&=&\sum_k b_k \psi_k=\sum_k \beta_k\psi^k,
\label{mirrosymmetry3}\end{eqnarray}
implying an infinite set of relations which we denote by
\begin{equation}
D(\alpha,\beta)=0.
\label{relations}\end{equation}
Now observe that performing the second quantization
\begin{equation}
\psi\to \hat \psi = \sum_j\left( \hat a_j\psi_j+\hat a_j^+
\overline\psi_j\right),
\label{secquant}\end{equation}
induces a quantization of the coefficients
$\alpha_j$'s. Therefore, whereas
the $\alpha_j$'s and $\beta_k$'s enter in
(\ref{relations}) as dual quantities,
in the second quantization the $\alpha_j$'s only become operators.
The important point is that Eq.(\ref{relations}), which is a manifestation
of the $x$--$\psi$ duality,
 suggests investigating
whether there exists a quantization with
the $\beta_k$'s considered as operators. Therefore it is natural to
consider
\begin{equation}
x\to \hat x=\sum_k( \hat\beta_k\psi^k+ \hat\beta_k^+\overline\psi^k).
\label{secquantb}\end{equation}
We now have two inequivalent dual pictures defined by
(\ref{secquant}) and (\ref{secquantb}) respectively.
Whereas Eq.(\ref{secquant}) corresponds to the second quantization
of the wave--function (associated to the Schr\"odinger equation
(\ref{erwin})),
Eq.(\ref{secquantb}) can be
considered as the quantization of the coordinate (associated to
Eq.(\ref{dualerwin}), dual to Eq.(\ref{erwin})).
We note that as
$\psi$ takes complex values we can use the notation
\begin{equation}
\hat x=\sum_k( \hat\beta_k z^k+ \hat\beta_k^+\bar z ^k).
\label{secquantbb}\end{equation}
This expression is conjectured in order to preserve the 
correspondence suggested by Eq.(\ref{mirrosymmetry3}).
The fact that this equation leads to the quantization of the 
coordinate should be further investigated.
In particular, we remark that the structure of Eq.(\ref{secquantbb})
resembles the expansion for the target coordinate in string theory.
For the time being, we note that inverting
$\psi=\psi(x)$ to $x=x(\psi)$
one obtains a description of geometrical quantities
in terms of the wave--function.
Therefore we can think at the inversion method as a way
to transfer quantum aspects directly to the coordinate, suggesting that
(\ref{secquantb}) should play a role in quantizing geometry.

We have seen that
both $\psi_E$ and ${\psi_E}_D$
enter on the same level, so that our formalism is manifestly
modular invariant.
An aspect which is related to this invariance is that
there are quantum structures which may be described in the
framework of monodromy transformations.
For example, by (\ref{jfwfd})
the $SL(2,{\bf Z})$ generators $S=\left(\begin{array}{c}0\\-1
\end{array}\begin{array}{cc}1\\0\end{array}\right)$
and $T=\left(\begin{array}{c}1\\0
\end{array}\begin{array}{cc}1\\1\end{array}\right)$
generate $|\psi_E^2|$ and $\psi_E^2$ respectively. These quantities
correspond to the probability densities
depending on if $\overline \psi_E\not \propto\psi_E$ or
$\overline \psi_E\propto\psi_E$.

A feature of our approach is that
it extends to higher dimensions. Furthermore,
it may also be applied to the case of the Klein--Gordon equation
(since the spinor components
satisfy the Klein--Gordon equation, the construction applies to the
fermionic case as well).

Let us first consider the
Schr\"odinger equation
\begin{equation}
\left(-{{\hbar^2}\over {2m}}\Delta+V(x)\right)\psi=E\psi,
\label{erwinD}\end{equation}
where $\Delta=\sum_{k=1}^{D-1}\partial_{x_k}^2$. The way to find the
generalization of Eq.(\ref{iu})
is to rewrite (\ref{erwinD}) in the form
\begin{equation}
\left(-{{\hbar^2}\over {2m}}\partial_{x_k}^2+V_k(x_k)\right)\psi=E\psi,
\label{erwinE}\end{equation}
for $k=1,...,D-1$, where we have introduced the ``effective potentials''
\begin{equation}
V_k(x_k)=\left[V(x)-{\hbar^2\over 2m\psi(x)}\sum_{j=1,j\ne k}^{D-1}
\partial^2_{x_j}\psi(x)\right]_{|x_{j\ne k}\; fixed}.
\end{equation}
Eq.(\ref{erwinE}) is now seen as a
second--order equation in the variable $x_k$, with
$x_{j\ne k}$ considered as parameters for the effective potential
$V_k$. Let $\psi_E^{(k)}$ and ${\psi^{(k)}_E}_D$ be linearly independent
solutions of Eq.(\ref{erwinE}).
Repeating the procedure considered in the one--dimensional case,
where now for any $k$ the
integration is taken from $x_{k0}$ to $x_k$ keeping the other
coordinate components fixed, we obtain
\begin{equation}
{\sqrt{2m}\over \hbar} x_k(\psi_E^{(k)})={\psi^{(k)}_E}^2{\partial
{\cal F}_E^{(k)}\over \partial ({\psi^{(k)}_E}^2)}-{\cal F}_E^{(k)},
\label{iubbb}\end{equation}
for $k=1,...,D-1$, and ($'\equiv \partial_{{\psi^{(k)}_E}}$)
\begin{equation}
4{{\cal F}_E^{(k)}}'''+(V_k(x_k)-E)\left(
{{\cal F}_E^{(k)}}'-\psi^{(k)}_E
{{\cal F}_E^{(k)}}''\right)^3=0,
\label{iocbbb}\end{equation}
which is an ODE for ${\cal F}_E^{(k)}(\psi^{(k)}_E)$ once $x_k$ in $V_k$
is replaced
with its functional dependence on $\psi^{(k)}_E$ given in
(\ref{iubbb}).
It is worth noticing that in the important case
$V(x)=\sum_{j=1}^{D-1} f_j(x_j)$, the functional structure of
${\cal F}_E^{(k)}$ does not depend on the ``parameters'' $x_{j\ne k}$.

In the case of the Klein--Gordon equation, we rewrite
$(\Box +m^2)\phi=0$ in the form
\begin{equation}
(\partial^{\mu}\partial_\mu+V_\mu(x)+m^2)\phi=0,
\label{ioiudhbb}\end{equation}
for $\mu=0,...,D-1$, where we have introduced the effective potentials
\begin{equation}
V_\mu(x_\mu)=\left[{1\over \phi(x)}\sum_{\nu=0,\nu \ne
\mu}^{D-1}\partial^{\nu}\partial_\nu
\phi(x)\right]_{|x_{\nu\ne \mu}\; fixed}.
\label{notevole}\end{equation}
The important difference with respect to the case of the Schr\"odinger
equation is that, as a consequence of its relativistic nature,
the time derivative appears in the Klein--Gordon equation at the
second--order.
This implies that the inversion formula also holds for the time component
$x^0=t$ and Eqs.(\ref{iubbb},\ref{iocbbb}) extend to the relativistic
case with $k\in [1,D-1]$ replaced by $\mu\in [0,D-1]$.

Another manifestation of the statistical structure underlying the
formalism is suggested by an analogy with $N=2$ SYM. The point is that,
in analogy with the role played by the scale $\Lambda$ in
Seiberg--Witten theory, we can interpret $\hbar$ as a parameter
defining the scale of a statistical system. In particular,
following the approach introduced in \cite{bonelli}, we first note that
for dimensional reasons
\begin{equation}
{K x\over \hbar}={\cal G}(\tau),
\label{3}\end{equation}
where $K=\sqrt{2mE}$, and
\begin{equation}
\tau = {\partial^2 {\cal F}_E\over \partial \psi_E^2}.
\label{4}\end{equation}
In this framework it makes sense to apply the operator
\begin{equation}
\hbar\partial_{\hbar},
\label{fundamental}\end{equation}
to Eq.(\ref{3}). We have
\begin{equation}
\beta \partial_\tau {\cal G}(\tau)=-{Kx\over \hbar},
\label{scaling}\end{equation}
where
\begin{equation}
\beta(\tau)=\hbar \partial_{\hbar}\tau.
\label{beta3}\end{equation}
Integrating (\ref{scaling}) we obtain
\begin{equation}
x={\hbar\over \hbar_0}
x_0e^{-\int^{\tau}_{\tau_0}dy{\beta^{-1}(y)}},
\label{fundamental2}\end{equation}
showing that the space coordinate has an anomalous dimension
determined by the ``beta--function'' (\ref{beta3}).
 In this context we observe that the
Heisenberg uncertainty principle depends on the scale
\begin{equation}
\Delta x \Delta p\ge \hbar=\hbar_0+ corrections.
\label{newHeisenberg}\end{equation}
We note that generalizations of the Heisenberg uncertainty principle
have been discussed in the context of different approaches
to quantum gravity in \cite{Amati}.

We observe that our approach sheds new light on the role
of the dual wave--function $\psi_{D_E}$. In this context
${\cal F}_E$ plays the crucial role as it can be seen as the
analogous of the Hamilton principal function. Actually, 
$\psi_E$ and ${\psi_D}_E=\partial{\cal F}_E/\partial\psi_E$ play
a similar role to $x$ and $p$ in Hamilton--Jacobi theory. 
The inversion formula Eq.(\ref{iu1}) is the key starting point for
this investigation \cite{FM}. 

Other aspects which merit further investigation
concern the possible role of our construction
in the framework of the stochastic approach to quantum
mechanics \cite{Guerra}, the
many--particle systems,
the case of coherent states and geometric quantization.
Here we limit ourself to observe that in the case of two--particle systems
with central potential $V(r)$, one can find
for $r$ an expression similar to (\ref{iu1}) with
$\psi_E$ and ${\psi_E}_D$ replaced by $\chi_E$ and
${\chi_E}_D=\partial_{\chi_E}{\cal
F}_{E}$
respectively  (we are using the standard notation
$\psi_E=Y_{lm}(\theta,\varphi)\chi_E(r)/r$).
 This is a consequence of the fact that both $\chi_E$ and
${\chi_E}_D$ are in the kernel of the operator
$\partial_r^2+2m(E-V)/\hbar^2-{l(l+1)/ r^2}$,
where $m=m_1m_2/(m_1+m_2)$ is the reduced mass.

In conclusion, we observe that starting from the inversion formula
we arrived at a statistical interpretation of the space coordinate
opening the way for a
possible understanding of the link between space--time structure
and quantum theory. In particular, we stress that the inversion
formula allows us to use
\begin{equation}
dx={\partial x\over \partial \psi}d\psi,
\label{interpret}\end{equation}
for connecting geometrical and quantum concepts.

It is well--known that after a Wick rotation of the time coordinate, the
 path--integral formulation of quantum mechanics resembles
the partition function of a thermodynamical system.
The appearance of the Legendre transform relating quantum and
macroscopic quantities may clarify this relationship suggesting a possible
thermodynamical interpretation of quantum mechanics whose
 implications
may bring about a new deep understanding of the
fundamental connection between geometry and quantum mechanics.

We thank G. Bonelli and M. Tonin for
important discussions and the
CERN Theory Division for hospitality.
This work was supported in part by DOE Grant No.\ DE-FG-0586ER40272 (AEF)
and by European Community Research Programme,
{\it Gauge Theories, applied supersymmetry and quantum gravity},
Contract SC1-CT92-0789 (MM).


\bibliographystyle{unsrt}

\end{document}